  \documentstyle[aps,multicol,psfig]{revtex}
\def\LamF{{\lambda_{\rm F}}}
\def\Kf{{k_{\rm F}}}
\def\Ef{{k_{\rm F}^{2}}}
\def\Hand{\hat{H}}
\def\amp{{\cal A}}
\newcommand{\eqbreak}{
\end{multicols}
\widetext
\noindent
\rule{.48\linewidth}{.1mm}\rule{.1mm}{.1cm}
}
\newcommand{\eqresume}{
\noindent
\rule{.52\linewidth}{.0mm}\rule[-.1cm]{.1mm}{.1cm}\rule{.48\linewidth}{.1mm}
\begin{multicols}{2}
\narrowtext
}
\begin{document} 
\draft 
\title{Low-energy quasiparticle states 
near extended scatterers in d-wave\\
superconductors and their connection with 
SUSY quantum mechanics}
\author{\.{I}nan\c{c} Adagideli, 
Paul M. Goldbart, 
Alexander Shnirman and Ali Yazdani} 
\address{Department of Physics and Materials Research Laboratory,   
University of Illinois at Urbana-Champaign, Urbana, IL 61801}
  \date{June 25, 1999}	
\maketitle
\begin{abstract} 
Low-energy quasiparticle states, arising from scattering 
by single-particle potentials in d-wave superconductors, are 
addressed.  Via a natural extension of the Andreev approximation, 
the idea that sign-variations in the superconducting pair-potential 
lead to such states is extended beyond its original setting of 
boundary scattering to the broader context of scattering by general 
single-particle potentials, such as those due to impurities.  The 
index-theoretic origin of these states is exhibited via a simple 
connection with Witten's supersymmetric quantum-mechanical model. 
\end{abstract}
\begin{multicols}{2}
\narrowtext
\noindent
{\sl Introduction\/}: 
In the present work we shall explore the low-energy quasiparticle 
states available in d-wave superconductors due to the presence of 
an extended scatterer such as a boundary or an impurity more than 
a few Fermi wavelengths across. In the context of boundary 
scattering, such states represent an important signature of 
sign-variations of the superconducting order parameter, as they 
have been shown to originate in the possibility of scattering 
between momentum orientations that are subject to superconducting 
pair-potentials of differing sign.  The main aims of our work are 
to extend the idea that sign-variations in the superconducting 
pair-potential lead to low-energy quasiparticle states to the 
context of scattering by general single-particle potentials, such 
as those due to impurities (i.e., beyond scattering by boundaries),
and to explore the robustness of this effect.  

The theoretical framework that we shall adopt is the semiclassical 
approach to the quantum-mechanical problem of scattering from the 
single-particle potential, via which the eigenvalue problem at hand 
reduces to a family of effectively one-dimensional problems for the 
particle-hole dynamics in the presence of the superconducting 
pair-potential.  Through this approach, we shall be able to see that 
the density of low-energy quasiparticle states (DOS) is determined 
solely by the {\it classical scattering properties\/} of the 
single-particle potential and, furthermore, that this DOS is 
insensitive to any suppression of the pair-potential that the 
impurity might cause.  This approach also provides us with a 
framework for classifying and calculating corrections to the DOS 
at low energies, such as those due to diffraction during scattering 
from the single-particle potential itself, or due to any 
pair-potential modifications beyond mere suppression 
(such as the induction of any out-of-phase components of the 
pair-potential).

Along the way, we shall discuss the fact that the emerging 
one-dimensional eigenproblem is a realization of Witten's 
supersymmetric quantum-mechanical model~\cite{REF:Witten81,REF:Junker}  
which, via the Witten index~\cite{REF:Witten81,REF:Junker},
provides a natural setting in which to explore zero-energy 
states~\cite{REF:Volovik,REF:IndexNotion}.  
Through this identification with 
Witten's model we shall see that the conditions under which 
zero-energy states exist are indeed those mentioned above, viz., 
propagation between pair-potentials of differing signs. In addition, we 
shall examine the role played by the semiclassical approximation to the 
scattering problem vis-\`a-vis the existence of zero-energy states, and 
thus see how it is that going beyond this semiclassical approximation 
generically introduces transition amplitudes between classical 
scattering trajectories, thus causing the dispersion of the formerly 
zero-energy states, e.g., into one or more low-energy peaks in the DOS.

We would like to stress at the outset that the issue of the origin of 
the low-energy states, viz., sign changes in the pair-potential, has 
already been soundly understood and extensively developed theoretically 
in several contexts: notable examples include the works of 
Buchholtz and Zwicknagl~\cite{REF:BuchBW} 
on p-wave superconductors near surfaces; 
and of Hu~\cite{REF:Hu}, 
Buchholtz et al.~\cite{REF:Sauls}, 
and Fogelstr\"om et al.~\cite{REF:Fogel97} 
on d-wave superconductors near flat surfaces.  
Low-energy states have also received extensive experimental 
attention in the context of boundary-scattering in 
high-temperature superconductors.  In particular, measurements of the 
(macroscopic) tunneling conductance~\cite{REF:Greene} have revealed a 
zero-bias anomaly indicative of the existence of low-energy states 
near boundaries. 

Apart from the effects of flat boundaries, theoretical research on  
low-energy quasiparticle resonances in d-wave materials has mostly been 
concerned with the effects of {\it point-like impurities\/} (i.e., 
impurities for which the size of the impurity is not much larger than 
the Fermi wavelength $\LamF$).  Of particular interest has been the 
effect of the impurity strength on the energies and wave functions of 
the resonances~\cite{REF:Balatsky1,REF:Balatsky2}.  More recently, 
attention has been paid to the effects on these resonances of 
impurity-induced suppression of the superconducting order 
parameter~\cite{REF:Hettler,REF:Sasha}.  Emerging from this body of 
work is a picture in which each strong, point-like impurity gives rise 
to a low-energy resonance.  This resonance, which would show up in 
the tunneling DOS as a pair of peaks symmetrically located around 
zero energy, transforms (in the particle-hole symmetric case) into 
a single, marginal, bound state at zero energy in the unitary scattering 
limit.  As the impurity strength is reduced, the energy of this 
resonance moves towards the gap maximum.  Moreover, the quantitative 
details of the band structure and/or order parameter can play 
important roles~\cite{REF:Flatte}. In particular, in particle-hole 
asymmetric systems the energies of the resonances no longer tend 
asymptotically to zero in the unitary limit. 

In contrast, the present work suggests that an extended (rather than 
point-like) impurity induces a zero-energy peak in the DOS with a weight 
of order the linear size of the impurity (measured in units of the Fermi 
wavelength).  Moreover, the resulting low-energy DOS is much less 
sensitive to details such as the precise form of the band structure 
and any in-phase order parameter variations, i.e., the peak at zero 
energy is inert.  In this respect, extended impurities behave more like 
flat boundaries than like point-like impurities. 

The theoretical distinctions between point-like and extended impurities
raised in this Letter have, to some extent, been addressed experimentally 
via scanning tunneling spectroscopy on ${\rm Bi_2 Sr_2 Ca Cu_2 O_8}$
surfaces~\cite{REF:JCDavis,REF:Yazdani}.  Work on native 
defects~\cite{REF:JCDavis,REF:Yazdani}, which often appear to be  
essentially point-like in STM imaging, yield weak signatures in the 
(smeared, local) DOS near each defect.  Such signatures can each be 
interpreted as being induced by a point-like impurity that yields a 
resonance of unit weight. In contrast, the artificially-induced defects 
described in Ref.~\cite{REF:Yazdani}, which appear to be more extended 
in STM imaging, show much stronger signatures in the DOS.  This is 
consistent with the idea that extended impurities produce many states, 
as the present work indicates they should. 

\noindent
{\sl Bogoliubov-de~Gennes eigenproblem\/}:
We regard the single-quasiparticle excitations as being described by the 
Bogoliubov-de~Gennes (BdG) 
eigenproblem~\cite{REF:AFAndreev,REF:Bruder}
\begin{equation}
\pmatrix{\hat{h} & \hat{\Delta} \cr
         \hat{\Delta}^{\dagger} & -\hat{h}}
\pmatrix{u\cr v}
=E\pmatrix{u\cr v},
\label{EQ:BdG}
\end{equation}
where the components 
$u({\bf x})$ and $v({\bf x})$ 
of the energy eigenstate respectively give the amplitudes for finding 
an electron and a hole at the position ${\bf x}$,  $E$ is the energy 
eigenvalue, and 
$\hat{h}=-\nabla^{2}-\Ef+V({\bf x})$ 
is the one-particle hamiltonian, 
in which $\Ef$ is the chemical potential 
[i.e., $\Kf$ ($\equiv 2\pi/\LamF$) is the Fermi wave vector]
and $V$ is the single-particle potential.  
We have adopted units in which $\hbar^2/2m=1$, 
where $m$ is the (effective) mass of the electrons and holes.
The operator $\hat{\Delta}$  (which should ultimately be 
determined self-consistently) is the pair-potential (integral) 
operator, whose action on the wave functions is 
specified by the (nonlocal) kernel 
$\Delta({\bf x},{\bf x}')$ via: 
$[\hat{\Delta}v]({\bf x})=\int d{\bf x}'
\Delta({\bf x},{\bf x}')\,v({\bf x}')$. 
We assume that sufficiently far from the scatterer 
$\Delta$ returns to the value that characterizes 
the bulk superconductor (e.g., s-wave, d-wave, mixed, etc.). 
As we shall see below, our computation of the low-energy DOS 
is insensitive to the precise form of any 
suppression of the superconducting order induced by the 
single-particle potential, and therefore continues to hold when 
$\Delta$ is replaced by 
its self-consistent value.  However, as we shall also see below, 
induced modifications of the superconducting order parameter that  
go beyond simple suppression in a manner that causes local 
supercurrents [i.e., via the addition of any intrinsically 
out-of-phase component to $\Delta$] spoil this robustness.

\noindent
{\sl Andreev's approximation for a strong single-particle potential\/}:
To analyze the BdG eigenproblem we first apply a 
semiclassical approximation, which reduces the full problem
to a family of first-order differential eigenproblems 
labeled by the classical trajectories of a particle 
at the Fermi energy in the presence of the full single-particle 
potential.  This amounts to extending the Andreev 
approximation to situations in which there is a 
single-particle potential whose energy scale $V_0$ is not negligible 
compared with the Fermi energy.  In technical terms, we 
are making an asymptotic approximation valid when $\Ef\gg(\Delta_{0},E)$, 
$V_0\sim\Ef$, and $V({\bf x})$ is slowly varying relative to $\LamF$.
To implement this approximation we consider the 
semiclassical solution of 
\begin{equation}
\big(
-\nabla^{2}-\Ef+V({\bf x})
\big)\,
\big(
\amp({\bf x})\, 
{\rm e}^{i\Kf S({\bf x})}
\big)=0, 
\label{EQ:SCwave}
\end{equation}
i.e., the \lq\lq large\rq\rq\ part of the BdG 
eigenproblem, where both $\amp({\bf x})$ and $S({\bf x})$
are taken to be slowly varying (with respect to 
$\LamF$)~\cite{REF:MaslovIndex}. 
By retaining the first and second powers in $\Kf$ we obtain, 
from Eq.~(\ref{EQ:SCwave}), the Hamilton-Jacobi equation
$\left|\bbox{\nabla}S({\bf x})\right|^{2}=
1-k_{\rm F}^{-2}V({\bf x})$ 
and the conservation condition
$\bbox{\nabla}\cdot
\left(
\amp({\bf x})^{2}\, 
\bbox{\nabla}S({\bf x})
\right)=0.$
We then use the resulting semiclassical solution, which is 
specified in terms of the incoming momentum orientation ${\bf n}$ 
via the asymptotic behavior 
$S({\bf x};{\bf n})\sim{\bf n}\cdot{\bf x}$~\cite{REF:MultiS} 
(for ${\bf x}$ far from the scattering center) and includes 
all of the fast (i.e., order of $\LamF$) variations of the exact 
BdG eigenfunctions, to perform a generalized 
separation of rapidly and slowly varying components by writing
\begin{equation}
\pmatrix{u({\bf x})\cr v({\bf x})}=
\amp({\bf x})\, 
{\rm e}^{i \Kf S({\bf x};{\bf n})} 
\pmatrix{\bar{u}({\bf x})\cr \bar{v}({\bf x})},
\label{EQ:sprtn}
\end{equation}
where $\bar{u}$ and $\bar{v}$ are assumed to be slowly varying 
relative to $\LamF$. Then, by inserting this form into 
Eq.~(\ref{EQ:BdG}) we obtain
\begin{eqnarray}
&&
\big[\hat{h}
\left(
\amp{\rm e}^{i \Kf S}\bar{u}
\right)\big]({\bf x})
\sim
-2i\Kf\,
\amp({\bf x})\,
{\rm e}^{i \Kf S({\bf x};{\bf n})}
\big({\bbox{\nabla}}S\big)
\cdot
\big({\bbox{\nabla}}\bar{u}\big),
\nonumber
\end{eqnarray}%
for the action of $\hat{h}$ on $\amp\,\exp\big(i\Kf S\big)\,\bar{u}$.

We now turn to the \lq\lq small\rq\rq\ part of the BdG eigenproblem, 
which involves the off-diagonal integral operator $\hat\Delta$.
It is convenient to transform to relative and center-of-mass 
coordinates, ${\bf r}$ and ${\bf R}$:  
\begin{equation}
\bar{\Delta}({\bf r},{\bf R})
\equiv\Delta({\bf x},{\bf x}'),
\quad 
{\bf r}\equiv{\bf x}-{\bf x}',
\quad
{\bf R}\equiv\frac{{\bf x}+{\bf x}'}{2}\,. 
\end{equation}
Then the action of $\hat\Delta$ can be asymptotically approximated 
(for $k_{\rm F}^{2}\gg\Delta_{0}$) as 
\eqbreak
\begin{mathletters}
\begin{eqnarray}
&&\left[\hat{\Delta}
\left(
\amp\,{\rm e}^{i\Kf S}\,\bar{u}
\right)\right]({\bf x})=
\int d{\bf r}\, 
\bar{\Delta}({\bf r},{\bf x-r}/2)\,
\bar{u}({\bf x-r}/2)\, 
\amp({\bf x-r}/2)\, 
{\rm e}^{i \Kf S({\bf x-r}/2;{\bf n})} 
\approx
\big(\amp({\bf x})\,
{\rm e}^{i\Kf S({\bf x};{\bf n})}\big)
\bar{u}({\bf x})\,
\Delta_{\rm eff}({\bf x};{\bf n}), 
\\
&&
\Delta_{\rm eff}({\bf x};{\bf n}) 
\equiv
\int d{\bf r}\, 
\bar{\Delta}({\bf r},{\bf x-r}/2)\,
\frac{\amp({\bf x-r}/2)}{\amp({\bf x})}\,
\exp\big(i\Kf S({\bf x}-{\bf r};{\bf n})
         -i\Kf S({\bf x};{\bf n})\big), 
\label{EQ:DelDef}
\end{eqnarray}%
\end{mathletters}%
\eqresume
\noindent
provided we assume that $\big(\bar{u}({\bf x}),\bar{v}({\bf x})\big)$ 
varies much more slowly than $\LamF$.  Thus the task of solving 
the full BdG eigenproblem~(\ref{EQ:BdG}) is reduced to the task of 
solving the (classical) Hamilton-Jacobi equation, along 
with the ($2\times 2$) first-order partial differential eigenproblem
\begin{equation}
\pmatrix{
-2i\Kf{\bbox{\nabla}}S\!\cdot\!{\bbox{\nabla}}&\!\!
\Delta_{\rm eff}({\bf x};{\bf n})\cr
\Delta_{\rm eff}^{*}({\bf x};{\bf n})&\!\!
2i\Kf{\bbox{\nabla}}S\!\cdot\!{\bbox{\nabla}}}
\pmatrix{
\!\bar{u}\!\cr
\!\bar{v}\!}
=E
\pmatrix{
\!\bar{u}\!\cr
\!\bar{v}\!}.
\label{EQ:1stOBdG}
\end{equation}
In fact, the eigenproblem is an {\it ordinary\/} rather than 
{\it partial\/} one.  To see this, recall the element of 
Hamilton-Jacobi theory~\cite{REF:Goldstein} in which one establishes 
that the solution $S$ of the Hamilton-Jacobi equation 
(at least for classically 
allowed regions) is indeed the action computed along the classical 
trajectory ${\bf x}_{\rm c}(\cdot)$ that solves Newton's equation
$\Ef\,\partial_{s}^{2}\,{\bf x}_{\rm c}(s)=
-{\bbox{\nabla}}V({\bf x}_{\rm c})$ 
subject to the condition $|\partial_{s}\,{\bf x}_{c}(s)|\rightarrow 1$ 
as $s\rightarrow\pm\infty$ (so that the classical motion is at 
the Fermi energy).  Owing to this connection between $\bbox{\nabla}S$ 
and ${\dot{\bf x}}_{\rm c}$, Eq.~(\ref{EQ:1stOBdG}) can be rewritten 
as~\cite{REF:MaslovNote}
\[
\Hand
\pmatrix{\bar{u} \cr \bar{v}}= 
E\pmatrix{\bar{u} \cr \bar{v}}, 
\quad
\Hand\equiv 
\pmatrix{-2i\Kf\partial_{s}&
\Delta_{\rm eff}(s)\cr 
\Delta_{\rm eff}^{*}(s)
&2i\Kf\partial_{s}},
\]
where $\Delta_{\rm eff}(s)$ is defined to be 
$\Delta_{\rm eff}({\bf x}_{\rm c}(s);{\bf n})$. 
This family of first-order ordinary differential eigenproblems 
is parametrized by ${\bf n}$ and the impact parameter $b$, which 
uniquely specify the classical trajectory ${\bf x}_{\rm c}(\cdot)$ 
from amongst those having energy $\Ef$. 

\noindent
{\sl Zero-energy states\/}: 
To search for zero-energy states it is useful to reduce the 
eigenproblem via the following sequence of steps.  We apply the 
unitary transformation (in electron-hole space) 
$\hat{U}\equiv\frac{1}{\sqrt{2}}
\bigl({1\atop{i}}{\phantom{-}1\atop{-i}}\bigr)$,
under which 
\begin{mathletters}
\begin{eqnarray}
\Hand 
&\rightarrow&
\Hand'
\equiv
\hat{U}^{\dag}\,\Hand\, \hat{U}=
\pmatrix{0&
\hat{A}\cr
\hat{A}^{\dag}&0\cr},
\label{EQ:Square}
\\
\hat{A}&\equiv&
-2i\Kf\partial_{s}-i\Delta_{\rm eff}(s), 
\,
\hat{A}^{\dag}\equiv
-2i\Kf\partial_{s}+i\Delta_{\rm eff}(s).
\label{EQ:Annihilate}
\end{eqnarray}%
\end{mathletters}%
We emphasize that it is not possible to arrive at this structure 
for values of $\Delta_{\rm eff}$ that are intrinsically complex 
(i.e., cannot be made real by an elementary gauge transformation), 
as is the case, e.g., for supercurrent-carrying states. 
The virtue of the structure of Eqs.~(\ref{EQ:Square}) and 
(\ref{EQ:Annihilate}) is that it allows us to recognize that 
zero-energy eigenfunctions of 
${\Hand'}$ have the form 
$\bigl({\varphi_+\atop{0}}\bigr)$ or 
$\bigl({0\atop{\varphi_-}}\bigr)$, 
where the functions $\varphi_{\pm}$ obey
\begin{equation}
\left(2\Kf\partial_{s}\mp \Delta_{\rm eff}\right)\varphi_{\pm}=0, 
\end{equation}
provided they exist (i.e., are normalizable).  Owing to their 
first-order nature, these (zero-energy) eigenproblems may readily 
be integrated to give 
\begin{equation}
\varphi_{\pm}(s)\propto 
\exp\left(\pm (2k_{\rm F})^{-1} \int^{s}ds'\,\Delta_{\rm eff}(s')\right).
\end{equation} 
However, the ability to normalize $\varphi_{\pm}$, and therefore 
the existence of zero-energy eigenvalues, depends on the form of 
$\Delta_{\rm eff}$ via the limiting values 
$\Delta_{\pm}\equiv\lim_{s\to\infty}\Delta_{\rm eff}(\pm s)$ 
for a given semiclassical path ${\bf x}_{\rm c}(\cdot)$. 
Specifically, for semiclassical paths for which 
$\Delta_{+}\,\Delta_{-}$ is negative, one or other (but not both) 
of $\varphi_{\pm}$ is normalizable and, therefore, for such 
paths provide {\it precisely one\/} zero-energy eigenvalue.  On the 
other hand, for semiclassical paths for which 
$\Delta_{+}\,\Delta_{-}$ is positive, neither of
$\varphi_{\pm}$ is normalizable, and therefore such 
paths provide no zero-energy eigenvalues.

This diagnostic for when semiclassical paths lead to zero-energy 
states allows us to assemble the zero-energy contributions to the 
DOS.  If, for the sake of concreteness, we restrict 
our attention to two-dimensional systems then our 
approximation to the low-energy DOS has the form
\begin{equation}
\rho_{\rm SC}(E)=
\delta(E)\,
\frac{\Kf}{2\pi}
\int d{\bf n}\,db\,
\left(1-{\rm sgn}\,\Delta_{+}\,{\rm sgn}\,\Delta_{-}\right). 
\label{EQ:DOSnear0}
\end{equation} 
This formula should have corrections, which vanish as $E$ tends to 
zero, coming from the nodes in the gap of the homogeneous d-wave 
state, as well as suppression of the superconducting state near  
the impurity. 

Let us now highlight some features of Eq.~(\ref{EQ:DOSnear0}).
(i)~The evaluation of Eq.~(\ref{EQ:DOSnear0}) requires only 
knowledge of the classical scattering trajectories for $V$.
(ii)~The DOS peak is located at zero energy. 
Corrections to this result, owing {\it inter alia\/} to 
particle-hole asymmetry, are of relative order 
$\max\big(1/\Kf R,\Delta_0/\Ef\big)$ (where $R$ is the characteristic 
extent of the impurity). For small $\Delta_0/\Ef$ and extended 
impurities these corrections are small. 
(iii)~Only the asymptotic signs of $\Delta$ at the ends of the 
classical trajectories feature; the DOS is unchanged by deformations 
of the pair-potential, provided the asymptotic signs are 
preserved and no out-of-phase components are induced.  
(iv)~The degeneracy of the zero-energy level is of order $\Kf R$,
the constant of proportionality being dependent on the form 
of $V$.

\noindent
{\sl Connection with Witten's model of supersymmetric 
quantum mechanics and index theory\/}:
Having seen, within the context of an explicit computation, the emergence 
(or otherwise) or zero-energy states, we now discuss the structure that 
underlies this issue, namely index theory~\cite{REF:IndexTheory}.  The 
relevant aspect of index theory is Witten's index from Witten's 
model of supersymmetric quantum mechanics (SUSY QM).  The specific 
connection is as follows: 
${\Hand}^{\prime 2}$ (c.f.~our~\ref{EQ:Square}) is Witten's SUSY 
Hamiltonian;  
$\Delta_{\rm eff}$ (our~\ref{EQ:DelDef}) is Witten's SUSY potential;  
$A$ and $A^{\dag}$ (our~\ref{EQ:Annihilate}) are proportional to 
Witten's annihilation and creation operators. 
Indeed, the analysis leading from Eq.~(\ref{EQ:Square}) to the 
conditions for the existence of a zero-energy state, mirrors the 
(by now) standard SUSY QM analysis. 

In SUSY QM, an important tool is the Witten index, i.e., the number of 
zero-energy states of the form $\bigl({0\atop{\varphi_{-}}}\bigr)$ 
minus the number of the form $\bigl({\varphi_{+}\atop{0}}\bigr)$.  
If the Witten index is nonzero then there certainly are zero-energy 
states (i.e., SUSY is good; see, e.g., Ref.~\cite{REF:Junker}, Sec.~2.1). 
If the Witten index is zero then there may or may not be zero-energy 
states, as contributions may cancel.  In the present context, we are not 
{\it prima facie\/} concerned with the Witten index and its properties, 
but rather with ascertaining the number of zero-energy states.  
However, owing to the fact that there is at most {\it one\/} zero-energy 
state for any semiclassical trajectory (because the normalizability 
condition cannot be simultaneously satisfied by both 
$\varphi_{+}$ and $\varphi_{-}$) the (modulus of the) Witten index does 
indeed permit the counting of the zero-energy states.

\noindent
{\sl Discussion and outlook\/}:
The condition on the existence of zero-energy states, together 
with Eq.~(\ref{EQ:DelDef}), provide us with a way of calculating 
the DOS at low energies by a simple counting of the number of 
classical trajectories that start and end with different signs of 
the superconducting pair-potential [see Eq.~(\ref{EQ:DOSnear0})].
Thus, the DOS at low energies depends only on the {\it classical 
scattering properties of the single-particle potential\/}.  

As we have stressed earlier, this result is valid in the regime in 
which the single-particle potential is both spatially extended and 
strong and the pair-potential is much smaller than the Fermi energy.  
Before turning to a discussion (and classification) of the generic 
corrections to this result for the DOS, which arise upon the relaxation 
of these conditions, we remark that the foregoing approximation scheme 
and results also hold for spatially extended single-particle 
potentials that are {\it weaker\/} than the Fermi energy.  Moreover, 
in the regime $V_0\alt\Delta_0$ our results can be extended to the 
case of rapidly-varying single-particle potentials (such as are due 
to point-like impurities).  However, 
as the strength of the single-particle potential is diminished, 
the classical trajectories will tend towards straight lines and, 
hence, the number of trajectories that \lq\lq see\rq\rq\ different 
signs of the pair-potential will be reduced.  This will result in a 
corresponding decrease in the degeneracy of the zero-energy level, 
in accordance with formula~(\ref{EQ:DOSnear0}).  Indeed, for 
$V_0\alt\Delta_0$ the trajectories are essentially straight lines.  
Thus, there would be no zero-energy states, but additional resonances 
(due to the impurity) may arise if the pair-potential is suppressed. 

By contrast, in the regime $V_0\sim\Ef$ but $V({\bf x})$ rapidly 
varying (e.g., for strong, point-like impurities), the approximation 
scheme that enabled us to reduce the problem to a family of 
one-dimensional eigenproblems breaks down, due to the fact that the 
previously-neglected $\bbox{\nabla}\amp$ term 
becomes comparable to previously-retained $\bbox{\nabla}S$ term.  The 
former term introduces diffraction effects in the (quantum-mechanical) 
scattering from the single-particle potential, as well as tunneling 
through the classically-forbidden region.  These effects can be viewed 
as consequences of nonzero transition amplitudes between states 
associated with the classical trajectories, and would result in the 
dispersion of the previously-degenerate zero energy states.

Let us conclude by remarking that the presence of an 
impurity-induced subdominant component to the pair-potential, provided 
it is {\it in-phase\/} with the dominant component, would not change 
the picture presented here: specifically, formula~(\ref{EQ:DOSnear0}) 
would continue to hold.  On the other hand, if an out-of-phase 
component is induced (e.g., so that locally the state becomes d+is), 
this would cause the zero-energy peak in the DOS to split into two 
peaks of nonzero width~\cite{REF:Greene,REF:Fogel97}, symmetrically 
disposed about zero energy, the lineshapes depending on the full 
(rather than solely the asymptotic) details of the pair-potential. 
If the out-of-phase component is small then the resulting lineshape 
can be computed via perturbation theory.  

\noindent 
{\it Acknowledgments\/}: 
Useful discussions with A.~V.~Balatsky and M.~Stone 
are gratefully acknowledged. 
This work was supported by 
the Department of Energy, 
Award No.~DEFG02-96ER45439, 
and by the Fulbright Foundation (A.S.).
  
\end{multicols}

\begin{references}
\bibitem{REF:IndexNotion}
The idea that index theory~\cite{REF:IndexTheory} can be used 
to address the origin of low-energy states in unconventional 
superconductors has been mentioned by a number of 
workers~\cite{REF:BuchBW,REF:Sauls}.
\bibitem{REF:MaslovNote}
The physical results presented here do not depend on 
Maslov indices because any phase acquired as a particle 
moves beyond a conjugate point as an electron is 
cancelled as the Andreev-reflected hole passes backwards 
through that conjugate point.
\end{references}
\end{document}